# High Current Density Vertical Tunneling Transistors from Graphene/Highly-Doped Silicon Heterostructures

Yuan Liu, Jiming Sheng, Hao Wu, Qiyuan He, Hung-Chieh Cheng, Muhammad Imran Shakir, Yu Huang and Xiangfeng Duan[*]

Dr. Y. Liu, J. Sheng, H. Wu, H. Cheng, Prof. Y. Huang

Department of Materials Science and Engineering, University of California, Los Angeles, CA 90095, USA;

Dr. Q. He, Prof. X. Duan

Department of Chemistry and Biochemistry, University of California, Los Angeles, CA 90095, USA; E-mail: xduan@chem.ucla.edu

Prof. M. I. Sharkir

Sustainable Energy Technologies Center, College of Engineering, King Saud University, Riyadh 11421, Kingdom of Saudi Arabia.

**Abstract**: Graphene/silicon heterostructures have attracted tremendous interest as a new platform for diverse electronic and photonic devices such as barristors, solar cells, optical modulators, and chemical sensors. The studies to date largely focus on junctions between graphene and lightly-doped silicon, where a Schottky barrier is believed to dominate the carrier transport process. Here we report a systematic investigation of carrier transport across the heterojunctions formed between graphene and highly-doped silicon. By varying the silicon doping level and the measurement temperature, we show that the carrier transport across the graphene/$p^{++}$-Si heterojunction is dominated by tunneling effect through the native oxide. We further demonstrate that the tunneling current can be effectively modulated by the external gate electrical field, resulting in a vertical tunneling transistor. Benefited from the large density of states of highly doped silicon, our tunneling transistors can deliver a current density over 20 A/cm$^2$, about two orders of magnitude higher than previous graphene/insulator/graphene tunneling transistor at the same on/off ratio.

## 1. Introduction

Graphene has emerged as an exciting electronic material due to its unique electronic properties and atomically thin geometry.[1, 2] However, graphene has zero intrinsic band gap, thus can not be directly used as the active channel material for logic transistors with sufficient on/off current ratio. Alternatively, with a finite density of states, tunable work-function and optical transparency, graphene can function as a tunable contact to create a new generation of electronic and optoelectronic devices.[3-10] In particular, the graphene/silicon heterostructure devices have attracted considerable recent interest. A unique feature in these devices is that the tunable work function of graphene produces a tunable barrier height across the graphene-silicon interface, which makes the graphene/silicon junction a great platform for the investigation of interface transport mechanisms as well as diverse device applications such as barristors,[11] photo-detectors,[12-15] high-speed modulators, [16] solar cells, [17-21] and chemical sensors. [22, 23] However, the studies to date are largely limited to the heterostructures between graphene and low-doped silicon with a considerable Schottky barrier, and the thermionic emission is believed to dictate the carrier transport process. The tunneling mechanism of native oxide is still unexplored. [20,24]

Here we report a new type of device based on heterostructures formed between graphene and highly doped silicon. By varying the doping level in silicon and the measuring temperature, we can clearly distinguish the competing carrier transport mechanism between thermionic effect through depletion region and the tunneling effect through the native oxide. Furthermore, we show that the tunneling current across the heterostructure can be effectively modulated by external gate electric field due to the unique tunable work function of graphene, resulting in a vertical tunneling transistor. Benefited from the high carrier density in the highly doped silicon, the tunneling transistor could deliver a current density over 20 A/cm$^2$, which is about two orders of magnitude higher than previously reported graphene/insulator/graphene tunneling transistors with a similar on/off ratio. [3, 25-28] Alternatively, with the ultrathin silicon oxide as the gate dielectric and highly doped silicon as the gate, we observe a strong bias-induced gating effect within the graphene transistor. Our studies demonstrate that the interfacial native oxide plays a crucial role in governing the carrier transport in graphene-silicon heterostructures, and can enable the creation of a new graphene-based tunneling transistors.

## 2. Device fabrication

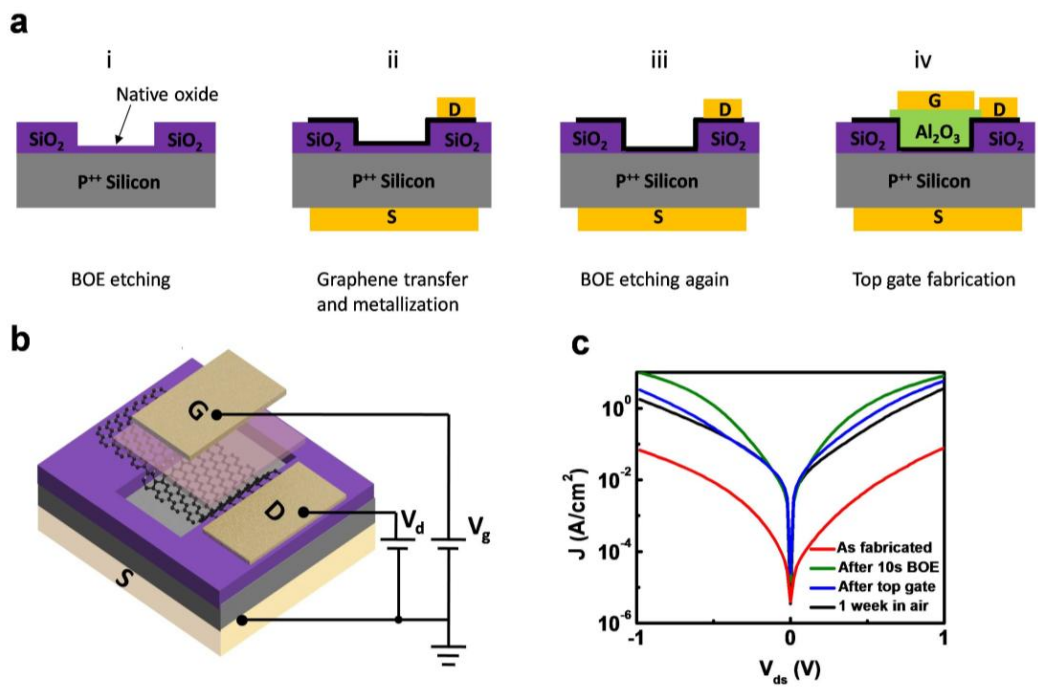

Figure 1. Device fabrication process and its schematics. a) Cross-section schematics of device fabrication process. (i), (ii), (iii), (iv) are four steps of the fabrication process in sequential order.

b) Perspective schematic illustration of the final device structure, where graphene is connected to the drain electrode (D) and highly doped silicon is used as the source electrode (S). c) Two terminal current after each step. The device current is relatively stable in ambient environment for 1 week after top gate fabrication, indicating that the gate dielectric and electrode could prevent further oxidation of silicon.

Figures 1a and 1b show the schematic fabrication process and the device structure. To fabricate the device, a 20 μm ×10 μm window was first defined on 300 nm thick silicon oxide via electron-beam lithography and etched using buffered oxide etcher (BOE), exposing the highly doped p-type silicon (~$10^{20}$/cm$^3$). The CVD-grown graphene[29] is then transferred onto the silicon surface using standard wet transfer technique. Next, graphene is etched into 10 μm wide stripes and contacted with drain electrode (D) (Cr/Au: 20/50 nm). The source electrode (S) is achieved by Ohmic indium contact on the back-side of the highly doped silicon substrate. The effective device area is defined by the overlapping junction area between graphene and silicon, and is 10 μm ×10 μm. This value is used to calculate the current density throughout the following sections.

It is well known that a thin layer of native oxide grows on silicon when exposed to the ambient conditions. In our experiment, the native oxide thicknesses were measured by ellipsometer (Gaertner Scientific, +/- 0.3 nm) to be 0.78 nm on fresh silicon, consistent with previous literature[20, 30]. However, because contamination of graphene with water or residual impurities during the transfer is usually inevitable, the native oxide layer between graphene and silicon can grow further. To investigate the role of the native oxide in device characteristics, a sequence of electrical measurement is performed during the device fabrication process. First, the as-fabricated two-terminal device was immediately measured in vacuum environment (Fig. 1a (ii)). A relatively small current density around 0.1 A/ cm$^2$ is observed across the graphene/silicon junction at 1 V bias (red line in Fig. 1c) due to the native oxide formed during the device fabrication process. Accordingly, we introduced a post-fabrication BOE treatment step to reduce the oxide between graphene and Si by dipping the device into BOE for another 20 s. The BOE solution can penetrate through the defective sites in graphene and etch the oxide layer between graphene and silicon[21] (Fig 1a (iii)), resulting in more than two orders of magnitude increase in current across the graphene-Si junction (green line in Fig. 1c). Finally, after the BOE treatment, we immediately deposit 20 nm Al$_2$O$_3$ as gate dielectric and Cr/Au (20/50 nm) as gate electrode. The source drain current decreased slightly after the gate fabrication process (blue lines, Fig. 1c), and remains relatively stable with additional one-week exposure in the air (black line, Fig. 1c), indicating that the gate dielectric and metal electrode can considerably slow down the further oxidation of silicon.

## 3. Results and discussion

The temperature dependent electrical transport properties of the final devices are measured using Physical Property Measurement System (PPMS, Quantum Design, Inc.). We first probed the two-terminal output curve without gate voltage, where graphene is used as the drain (D) electrode and the highly doped silicon is grounded as the source (S) electrode. The graphene/Si junction displays a non-linear I-V curve (Fig. 1C and Fig. S1), indicating an energy barrier in the charge transport process. In general, two mechanisms could contribute to the barrier between graphene and semiconductor: the Schottky barrier for thermionic emission, and tunneling barrier across the native oxide. [31] The relative contributions of these two mechanism depend on both the measurement temperature and the doping level of silicon. A standard criterion can be set by comparing the thermal energy (kT) to the characteristic energy parameter ($E_{00}$) defined as[31-33]

$$E_{00} = \frac{qh}{4\pi}\sqrt{\frac{N}{m^*\varepsilon_s}} \quad (1)$$

where N is the carrier concentration of silicon, $m^*$ is the effective mass, h is the Plank constant and $\varepsilon_s$ is the permittivity of silicon. When $E_{00}$< kT, thermionic emission dominates carrier transport; and when $E_{00}$> kT, the tunneling process dominates the charge transport. In our case, the highly doped silicon with a carrier concentration around $10^{20}$/cm$^3$ is used, resulting in an $E_{00}$~0.1 eV according to equation 1. This value is larger than kT even at room temperature ($kT_{300} \approx 26$ meV), indicating that the tunneling current dominate the transport throughout the temperature range from 300 K to cryogenic temperature. This is further supported by the temperature dependent measurements. Figure 2a shows the current as a function of temperature at various bias voltages, with the I-V curves at different temperatures shown in Fig. S1. We found that the current is insensitive to the temperature at all bias voltages, confirming the carrier transport is dominated by the tunneling process. The tunneling barrier here consists of primarily native silicon oxide (~1 nm) and a thin depletion layer in silicon (<1 nm), and will be discussed in detail.

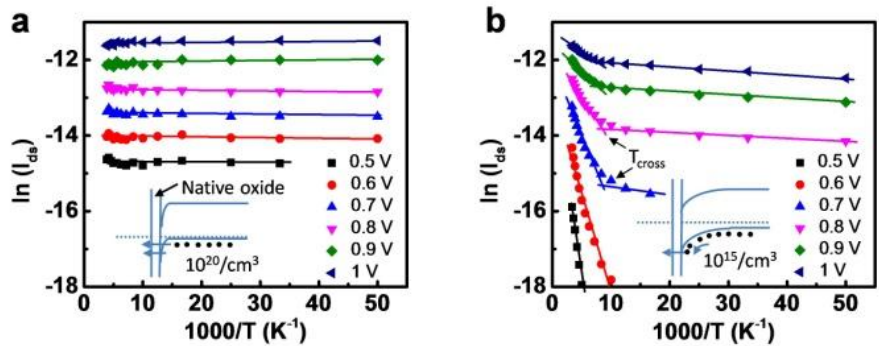

Figure 2. Transport characteristics of graphene/silicon junction with highly doped and lightly doped silicon. a) $I_{ds}$ at different bias voltage as a function of temperature for graphene/highly-doped-silicon (~$10^{20}$/cm$^3$) junction. Inset is the schematic band diagram. Due to thin depletion region, positive carriers can directly tunnel through the depletion region and native oxide, resulting in a tunneling current insensitive to the temperature. b) $I_{ds}$ at different bias voltage as a function of temperature for graphene/lightly-doped-silicon (~$10^{15}$/cm$^3$) junction. Inset is the schematic band diagram. Due to relatively thick depletion region, positive carriers need to be thermally activated through the depletion region and then tunnel across the native oxide, showing a clear $T_{cross}$ in the I-T curve, with different dominating transport mechanism at different temperature.

To better understand the competing relationship between native oxide and depletion region, we have also fabricated the control sample on lightly doped silicon substrate, where 8 μm thick p-type (~$10^{15}$/cm$^3$) silicon is epitaxially grown on p$^{++}$ substrate. The epitaxial wafer with highly doped substrate is used here to reduce the series body and contact resistance from the substrate. The temperature dependent measurements (Fig. 2b) of the resulting device show a clear crossover ($T_{cross}$) from an exponential temperature dependence at high temperature regime to a very weak temperature dependence at low temperature, suggesting two different dominating transport mechanisms at different temperature regime. With lightly doped silicon, the depletion region is much wider (~700 nm in this case) and the tunneling probability is greatly reduced. Thus, at high temperature > $T_{cross}$, the carriers are thermally activated to cross this depletion barrier and subsequently tunnel through the thin native oxide layer (Fig. 2b, inset). As a result, the source-drain current decreases exponentially with the temperature in this regime. Upon reduction of temperature below $T_{cross}$, the carriers do not have enough thermal energy to overcome the Schottky barrier anymore. At this point, carrier transport is dominated by the tunneling process through depletion region and the native oxide, with much lower current and little temperature dependence.

Next, we examined the gate modulation effect on the tunneling current in graphene/highly-doped silicon heterostructure. The output characteristics at different gate voltages clearly show an increase of source-drain current with decreasing gate voltage towards negative direction (Figure 3c), mimicking a p-type transistor behavior. This is consistent with p-type silicon used here, with the holes as the majority tunneling carriers. The tunneling device showed a current density of 20 A/cm$^2$ at the bias of 1 V, which is about 2 orders of magnitude larger than the previously reported graphene/ insulator/graphene devices at similar on/off ratio[3, 25-28]. The large current density can be attributed to the p$^{++}$-silicon used here, which has a larger density of states compared with graphene used previously, leading to a greatly increased tunneling probability (see the equation 4 in the discussion below). It should be noted that the achievement of a large current density is central to the performance of a tunneling transistor because the intrinsic delay of a transistor ($\tau = CV/I$) is inversely proportional to the deliverable current density. Moreover, the on/off ratio in our device is typically 50 to 100 under 1 V bias (Fig. 3c, 3d), comparable with previous reported vertical tunneling devices. Further increasing the bias voltage will increase the current density, yet at the sacrifice of the on-off ratio. At the bias voltage of 2 V, the $I_{ds}$-$V_g$ curve shows a maximum current density over 100 A/cm$^2$ with a smaller on-off ratio less than 10.

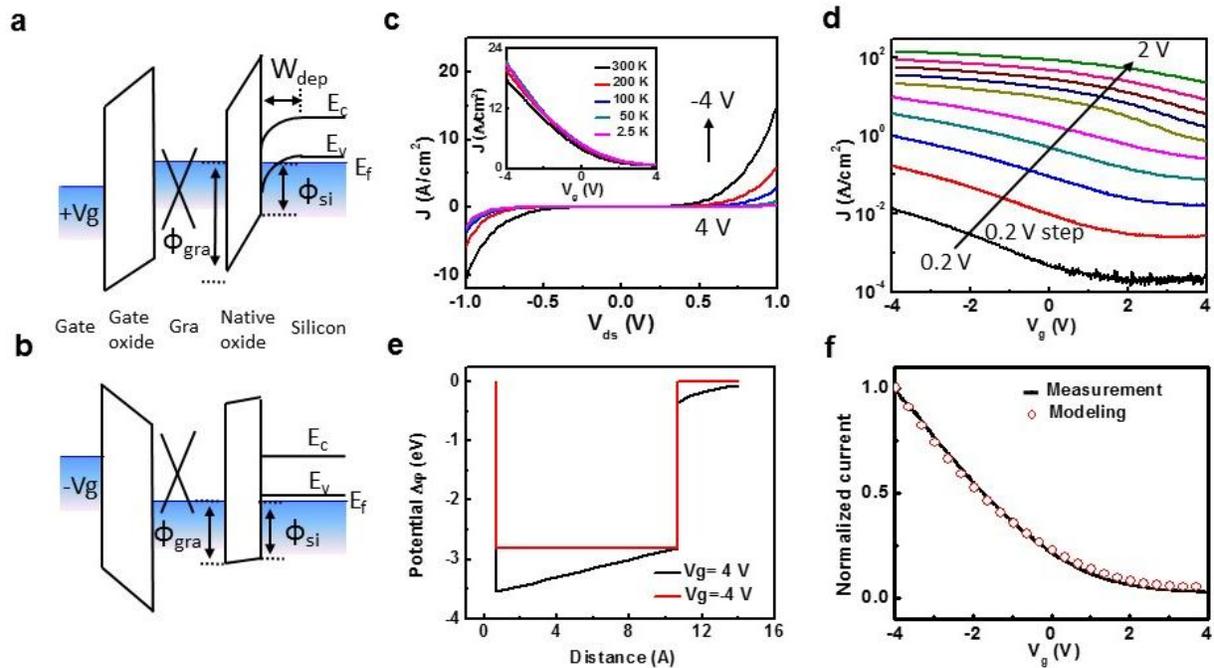

Figure 3. Gate modulation of the vertical tunneling current. a), b), Band diagram of graphene-highly doped silicon junction. Under positive gate voltage (a), graphene is n-doped and silicon surface is depleted, resulting in a larger tunneling barrier. Under negative gate voltage (b), graphene is p-doped and there is no surface depletion of silicon, making the barrier smaller and the device is under ON state. c) $I_{ds}$-$V_{ds}$ output characteristics under different gate biases from -4 V to 4 V. Inset shows its transfer curve at different temperatures and the bias voltage is 1 V. d) Semilog plot of the transfer curve at different bias voltages. Large on-off ratio around 50~100 can be achieved at small bias voltage (<1 V). Increasing the bias voltage will increase the current density at the sacrifice of on-off ratio. e) Calculated hole tunneling barrier relative to Fermi level of graphene at ON (-4 V, red line) and OFF (4 V, black line) states. f) Calculated tunneling

current as a function of gate voltage (red dot), which is consistent with measurement data (black line). The bias voltage is 0.5 V.

In general, the gate voltage applied across the vertical junction can modulate the work function of the graphene (Fig. 3a, b), as well as the charge concentration on the silicon surface (depletion/accumulation) and therefore the relative tunneling barrier across the graphene and silicon junction.[34] As shown in schematic band diagram (Fig. 3a), a positive gate voltage reduces the work function of graphene, thus increases barrier between graphene and $SiO_2$ ($\varphi_{gra}$) for hole transport. Additionally, a positive gate voltage could deplete p-type silicon surface, resulting in a depletion layer as an additional tunneling barrier (Fig. 3a and 3e). Together, these two effects suppress the tunneling probability and reduce the amplitude of the tunneling current to result in an OFF state (Fig. 3c, purple curve). Sweeping gate voltage toward the negative direction increases the work function of graphene, which in turn reduces the hole tunneling barrier height with the silicon oxide ($\varphi_{gra}$). Additionally, with a negative gate voltage, the silicon surface is also switched to the accumulation region without additional barriers. Together, the negative gate voltage can greatly enhance the tunneling current and switch the device into an ON state (Fig. 3c, black curve). Moreover, n-type vertical tunneling transistor with the opposite transfer behavior is also observed in graphene/n$^{++}$-Si heterostructure (Fig. S2), where the polarity of the transfer curves is reversed because the electrons become the majority tunneling carriers.

To quantitatively analyze the gate-dependent tunneling current in our graphene/p$^{++}$-Si junction, we have modeled our structure using Gauss's Law to account for the charge balance in the vertical structure (Supplementary information 3). With this approach, the tunneling barrier height under zero bias can be calculated and plotted (Fig. 3e). At a positive gate voltage of 4 V, the barrier height at the graphene side ($\varphi_{gra}$) is around 3.44 eV and that at the silicon side ($\varphi_{si}$) is around 2.96 eV. The depletion width of the highly doped silicon is around 0.8 nm according to the equation:

$$W_{dep} = \frac{\varepsilon_{SiO_2}}{eN} F_b, \qquad (2)$$

This depletion layer result in an additional parabola shaped tunneling barrier. Here $\varepsilon$ is the permittivity of SiO$_2$ and $F_b$ is the electric field across the native oxide. At the negative gate voltage of -4 V (Fig. 3e, red curve), the graphene is p-doped, resulting a smaller work function ($\varphi_{gra}$~2.91 eV). Additionally, the surface of silicon is under accumulation without band bending and the $\varphi_{si}$ is 2.8 eV. After quantitatively analyzing the tunneling band structure, the tunneling transmission coefficient can be modeled using WKB approximation (see Supplemental information 4) according to the equation 3:

$$T=\exp\left(-2\int_0^d dx \sqrt{\frac{2m^*}{\hbar^2}\Delta(x)}\right) \quad (3)$$

where $m^*$ is the effective electron mass, $\hbar$ is the reduced Plank constant, d is the thickness of native oxide plus depletion region width, and $\Delta(x)$ is the hole barrier height as plotted in Figure 3e. Using the calculated tunneling coefficient, we can model our tunneling current according to

$$I \propto \int dE\, DOS_{Si}(E-eV)\, DOS_{gra}(E)\, T(E)\, [f(E-eV)-f(E)] \quad (4)$$

where the $DOS_{Si}$, $DOS_{gra}$ refer to the density of states of silicon and graphene, respectively. $T$ is the transmission coefficient through the tunneling barrier, and $f$ is the Fermi distribution function (see Supplementary information 4). The modeling results are plotted as the red dots in figure 3f, which show a high degree of consistency with our experimental data (black curve).

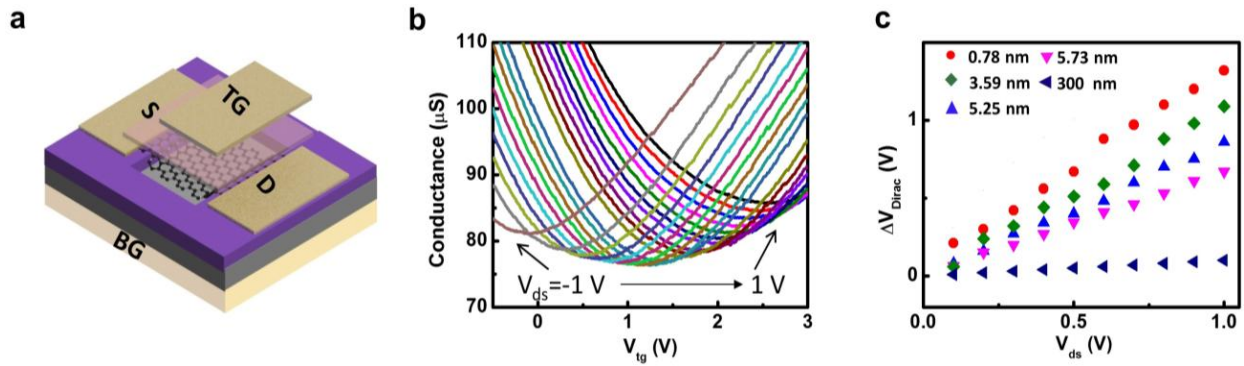

Figure 4. Dual-gate graphene transistor with different oxide thickness. a) Device schematics are shown, where graphene is connected to source, drain electrodes, highly doped silicon works as back gate and thin oxide is used as back gate dielectric. b) Transfer curve of graphene under different bias voltages ($V_{ds}$) from -1 V to 1V. It is clear the Dirac point voltage shifts

significantly with the bias voltages, due to a strong bias-induced gating effect. c) The shifted Dirac point as a function of bias voltage, where the back oxide thickness can be derived from the slope. Five different thicknesses (0.78 nm to 300 nm) are measured from ellipsometer before graphene transfer.

To further understand the role of native silicon oxide in the graphene/ $p^{++}$-silicon junction and determine native oxide thickness, we have measured the current transport across the graphene channel with the device connection shown in Figure 4a. In this case, both source and drain electrodes are connected to graphene. The top-gate electrode and the Si substrate are used as a local gate and global back-gate to control the carrier concentration and polarity in the graphene layer. Figure 4b shows the conductance of graphene device as a function of top-gate voltage measured at different source-drain bias from -1 V to 1 V, and $V_{bg}$=0 V (Si grounded). We found that the Dirac point's position of graphene is strongly dependent on the source-drain bias, indicating a large drain-induced gating effect. At $V_{ds}$=-1 V, the whole channel has negative potential, where the grounded silicon has relative higher potential than the channel with $V_{bg-drain}$>0. At this point, the channel is n-doped by back gate and the Dirac point is shifted to negative (-3.17 V) direction. The opposite is true for $V_{ds}$>0, where the Dirac point is shifted to 0.94 V at $V_{ds}$=1 V. It should be noted that the observed bias induced gating effect is much stronger than that reported previously,[35] due to the ultra-thin native oxide (~1 nm) as back gate dielectric. In order to confirm this, we have fabricated the same device with different oxide thickness. The oxide is thermally grown intentionally before graphene is transferred and the thickness is measured using ellipsometer. The bias induced gating effects for the devices with different oxide thickness are summarized in Figure 4c. It is apparent that this effect is negligible on 300 nm thick $SiO_2$ and becoming more pronounced with thinner silicon oxide (Supporting information 6).

## 4. Conclusion

In summary, we have presented a new vertical tunneling device structure based on graphene, native oxide and highly doped silicon. By using highly doped silicon, we demonstrate that the carrier transport across the graphene/Si heterostructure is dominated by tunneling effect through the native oxide. We further show that the tunneling current can be effectively modulated by the external gate electrical field, resulting in a vertical tunneling transistor. Benefited from large density of state in highly doped silicon, the tunneling transistor could deliver a current density over 20 A/cm$^2$ at 1 V bias, while retaining an on-off ratio up to 100. Our observation suggests that the doping concentration and interfacial native oxide are crucial in graphene-silicon hybrid system, which could lead to a new pathway to graphene-based tunneling transistors, as well as a better understanding of the tunneling barrier in two-dimensional vertical heterostructures.

**Acknowledgements**

We acknowledge the Nanoelectronics Research Facility (NRF) at UCLA for technical support. X.D. acknowledges partial support by ONR through grant number N00014-15-1-2368.

# High Current Density Vertical Tunneling Transistors from Graphene/Highly-Doped Silicon Heterostructures

Yuan Liu, Jiming Sheng, Hao Wu, Qiyuan He, Hung-Chieh Cheng, Muhammad Imran Shakir, Yu Huang and Xiangfeng Duan[*]

**Supplementary Information**

1. $I_{ds}$-$V_{ds}$ of graphene/ $p^{++}$ silicon junction with linear plot and semilog plot at various temperature.

2. $I_{ds}$-$V_{ds}$ output characteristic and $I_{ds}$-$V_g$ transfer curve of graphene/$n^{++}$ junction.

3. Simulation of graphene/$p^{++}$ tunneling barrier band structure.

4. Simulation of graphene/$p^{++}$ tunneling transistor possibility at various gate voltage and bias voltage.

5. Finite element method calculation of the bias-induced gating effect

6. Measurement of oxide thickness by dual-gate effect

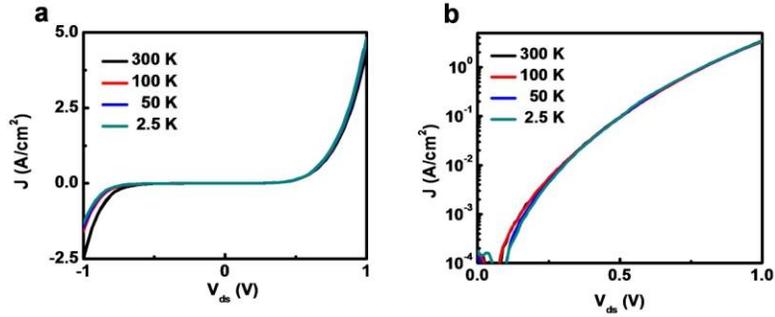

**Figure S1.** Output characteristic of graphene/p$^{++}$ junction at various temperature. Linear plot (**a**) and semilog plot (**b**) of I-V curve shows no major difference at temperature from 300 K to 2.5 K, indicating the tunneling mechanism.

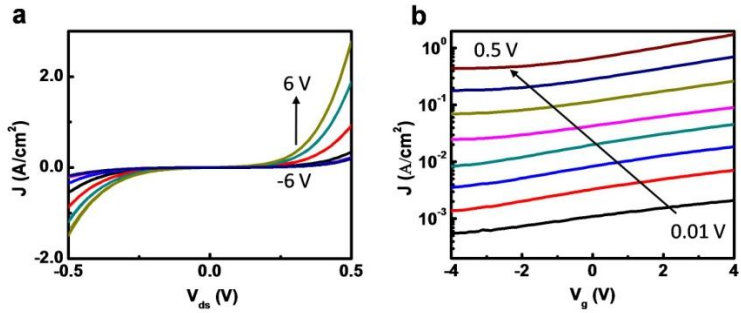

**Figure S2.** Electrical properties of graphene/n$^{++}$ tunneling transistor. **a**, I-V curve at different gate voltage from -6 V to 6 V with 2 V step. The current increase with gate voltage, mimicking a NMOS behavior. (**b**) Transfer curve of the same device shows an on-off ratio around 10 at different bias voltage. The bias voltage is 0.01 V, 0.02 V 0.05 V, 0.1 V, 0.2 V, 0.3 V, 0.4 V and 0.5 V, respectively.

### 3. Simulation of graphene/p$^{++}$ tunneling barrier band structure:

In order to decide the graphene/p$^{++}$ tunneling barrier band structure, first we calculate the Fermi energy, $E_f$, of the monolayer graphene relative to its Dirac point. When $E_f$>0, the graphene is n-doped, and vice versa. The Fermi level of highly doped silicon is assumed to coincide with valence band edge of Si and with conduction band edge for n$^{++}$ silicon[1]. The potential difference between graphene and gate electrode, $V_b$-$V_g$, contains contributions from 3 sources: first, the difference of work function ϕ in two materials, i.e. the flat band voltage; second, the potential drop caused by electric field $F_g$ in the dielectric $Al_2O_3$; finally, shift of the graphene's Fermi level relative to its Dirac point, (i.e. the quantum capacitance contribution). As a result, we have the following equation:

$$V_b\text{-}V_g =-(\phi_{gate} - \phi_{graphene})+F_g d_{Al_2O_3} - E_f \qquad (S1)$$

Here ϕ is the work function of the material in electron volt and $d_{Al_2O_3}$ is the thickness of the $Al_2O_3$ layer.

Similar argument can also be applied when writing down the potential difference between the graphene and ground if we also account for the depletion effect in Si. Take p$^{++}$ Si as an example. When $F_b$>0, net negative charges will be present at the interface between Si and $SiO_2$. This will lead to a depletion region of finite width across which there will also be a voltage drop[1]. The relation between the electric field $F_b$ in $SiO_2$ and the voltage drop $\phi_{depletion}$ across the depletion region is:

$$\phi_{depletion} = \begin{cases} \frac{\varepsilon_{SiO_2}}{2eN_{poly}} F_b^2, & F_b>0 \\ 0, & F_b<0 \end{cases}$$

for p$^{++}$ Si and

$$\phi_{depletion} = \begin{cases} \frac{\varepsilon_{SiO_2}}{2eN_{poly}} F_b^2, & F_b<0 \\ 0, & F_b>0 \end{cases}$$

for n++ Si. Here $\varepsilon_{SiO_2}$ is the dielectric constant of the $SiO_2$ and $N_{poly}$ is the dopant concentration in Si. Then S1 can be written as:

$$V_b - 0 = -(\phi_{ground} - \phi_{graphene}) + F_b d_{SiO_2} - E_f + \phi_{depletion} \quad (S2)$$

The electric field $F_g$ in $Al_2O_3$, and $F_b$ in $SiO_2$ can be related to the induced carrier density in graphene layer by Gauss's Theorem as follows:

$$\varepsilon_{SiO_2} F_b + \varepsilon_{Al_2O_3} F_g = \mp \lambda e \quad (S3)$$

$\lambda$ is the carrier concentration value (p or n). "+" sign is picked in positively doped case when $E_f < 0$ and "-" sign is picked with negative doping when $E_f > 0$.

The electronic property of graphene leads to the description of its carrier concentration $\lambda$ with $E_f$ as follows:

$$\lambda = \frac{E_F^2}{\pi \hbar^2 v_F^2} \quad (S4)$$

, which is a result of graphene's linear dispersion relation and 2-D nature of electrons[2-5]. Note that these properties of graphene also implies that its density of states is proportional to the state's energy relative to Dirac point, a conclusion that will be used later.

Combining equations (S1) through (S4) yields a quartic equation that allows us to solve for $E_f$ under given device geometry and $V_b$, $V_g$ values. This equation is solved by numerical means using MATLAB R2009a.

With $E_f$ solved, we can plug it into equation (S1) to get $F_b$. $E_f$ and $F_b$ then allow us to calculate depletion region's width (according to equation (2) in main text), barrier height, as well as the potential drop on the oxide and eventually complete the band diagram.

**4. Simulation of graphene/p++ tunneling transistor current at various gate voltage and bias voltage.**

From standard tunneling theory[6], the hole tunneling current I can be written as follows:

$$I \propto \int dE\, DoS_{poly-Si}(\text{E-eV})\, DoS_{graphene}(E)\, T(E)\, [f(E-eV)-f(E)] \quad (S5)$$

where DoS stands for density of states, T is the tunneling coefficient and f(E) is the Fermi-Dirac distribution function.

As aforementioned, $DoS_{graphene}(E)>0$ is proportional to the state's energy relative to Dirac point, i.e.

$$DoS_{graphene}(E) \propto |E - E(\text{Dirac point})|. \quad (S6)$$

Here we assume that: (i) $DoS_{Si}$ is a constant outside of its bandgap and shrinks to a smaller constant within the bandgap[7] (it is assumed to be zero). The ratio between two constants depends on the impurity energy levels within the bandgap and we used the value 0.2 in our simulation to give a best fit; (ii) T(E) can be calculated by WKB method[8,9] when we consider both the oxide and depletion region tunneling:

$$T(E) = T_0(E) T_{dep}(E),$$

where

$$T_0(E) = \exp\left(-2\int_0^d dx\, \sqrt{\frac{2m}{\hbar^2}\Delta(x)}\right) \quad (S7)$$

is the standard WKB formula for quantum tunneling. The integration is carried out through the thickness of $SiO_2$ and $\Delta(x)$ is the height of the trapezoidal barrier, which can be obtained based on the band diagram acquired in section S3.

$T_{dep}(E)$ refers to the contribution of depletion region tunneling. The depletion region is simulated as a parabola shaped barrier with a height of $\phi_{depletion}$ and $T_{dep}(E)$ is calculated in the same manner as in equation (S7) when there is depletion. In the case there is no depletion $T_{dep}(E)$ is simply 1.

Temperature is assumed to be small enough so that f(E) can be taken as step-like, i.e.

$$f(E)=1 \text{ when } E<\mu \text{ and } f(E)=0 \text{ when } E>\mu \quad (S8)$$

Finally, tunneling current I can be calculated with $E_f$ solved in section S3 and equations (S5)-(S8).

## 5. Finite element method calculation of the bias-induced gating effect

In order to decide the potential distribution and the average potential in the channel, the graphene channel is treated 1-dimensionally and evenly divided into N elements. Each element acts as a resistance in the circuit model, whose value is modulated by the local gate voltage it "feels", i.e. the difference between back gate voltage $V_{bg}$ and the local potential where the element lies. These resistances connect the source and drain electrodes in a serial pattern. A larger N generally helps to increase calculation accuracy but also calculation time. Also, the length of each element should be large compared to graphene's lattice constant for the resistance model to be valid. N=1000 is used in the calculation and the length of each element is 10 μm/1000=10nm≈ 40*graphene's lattice constant.

The local potentials at the boundaries of these 1000 elements are denoted $V_1=V_s$ (source), $V_2$, $V_3$, …$V_{1000}$, and $V_{1001}$ and the resistances of these elements are denoted $R_1$, $R_2$, $R_3$, …$R_{1000}$. The current I is consistent throughout the circuit. As a result, we have

$$V_{i+1}\text{-}V_i=I*R_i \tag{S9}$$

where i runs from 1 to N.

The graphene elements' resistances, $R_i$, are determined by the local gate voltages, $V_{bg}$- $V_i$ (for short elements $V_i \approx V_{i+1}$). For our device, the relation between these two can be fitted from our measurement:

$$R_i=2.60407\text{e-}4+-5.52312\text{e-}4*(V_{bg}-V_i)+7.17065\text{e-}4*(V_{bg}-V_i)^2$$

$$+2.40889\text{e-}4*(V_{bg}-V_i)^3+-3.49274\text{e-}4*(V_{bg}-V_i)^4 \tag{S10}$$

(Note that since we are interested only in potential distributions, absolute unit for R or I is not essential).

Now with a given value of $V_{bg}$, $V_1=V_s$ and a trial solution I', the value of $V_{1001}$ can be calculated and compared with $V_d$. A new trial solution can be generated according to the comparison result. In this way we can solve for I recursively with desired accuracy and also the channel potential distribution as well as average potential (a uniform potential value throughout the channel if the same current is to be sustained).

When both the source and back gate is grounded, and $V_{ds}=1$V, the average channel potential is calculated to be 0.64 V.

## 6. Measurement of oxide thickness by dual-gate effect

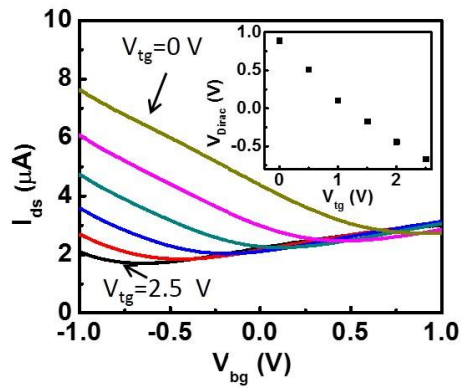

**Figure. S3**. Transfer curve of graphene by sweeping the back gate voltage, at different top gate voltage (0 V to 2.5 V). The Dirac point shift is plotted in the inset, where the back-gate oxide thickness can also be derived from the slope.

We can also quantitatively determine the oxide thickness using the bias induced gating effect described above. The average channel potential can be calculated through finite element method (FEM) (Supplemental information 5). When both the source and back gate is grounded, and $V_{ds}$=1V, the average channel potential is calculated to be 0.64 V. The silicon oxide thickness can be derived from the slope of Figure 4c using equation[10]:

$$t_{ox} = \frac{0.64 \times \varepsilon_{SiO2} V_{bias}}{C_{Al2O3}(V_{tg}-V_{Dirac})} \quad (S11)$$

where $\varepsilon_{SiO2}$ is the dielectric constant of silicon oxide, and $C_{Al2O3}$ is the capacitance of top gate dielectric (~270 nF/cm$^2$).

**Table S1. The oxide thickness by measurement and fitting**

| Measured (Ellipsometer) | Fitting (Bias gating effect) | Fitting (Dual gate effect) |
|---|---|---|
| 5.73 nm | 8.27 nm | 6.68 nm |
| 5.25 nm | 6.44 nm | 5.59 nm |
| 3.59 nm | 5.08 nm | 4.48 nm |

The calculated results are shown in table S1, which is largely consistent with the values measured by ellipsometer. The oxide thickness can be further studied by sweeping back gate voltage as a dual-gate transistor. As shown in Figure S3, increase the top gate voltage will result in a negative shift of Dirac point, consistent with previous report on dual-gate graphene transistor. The relationship of the Dirac point and top gate voltage is plotted in the inset (Fig. S3). The slope represents the ratio between the top-gate and back-gate capacitances $C_{tg}/C_{bg}$, where the back gate oxide thickness can be further derived from the slope according the equation[10]:

$$t_{ox} = \frac{\varepsilon_{SiO2} dV_{bg}}{C_{Al2o3} dV_{tg}}, \quad (S12)$$

The derived data is also summarized in Table S1. We note that the calculated oxide thickness from both effects (bias induced gating effect and dual-gate effect) are a litter higher than those determined by ellipsometer measurement. The difference could be due to the interlayer distance between graphene and silicon, as well as the effect of thin depletion region inside highly doped silicon.